\documentclass[useAMS,usenatbib]{mn2e}
\usepackage{amssymb}
\usepackage{graphicx}
\usepackage[usenames]{color}
\usepackage{url}

\newcommand{\de} {{\rm d}}

\begin{document}

\title{Compact formulae, dynamics and radiation of charged particles under synchro-curvature losses}

\author[D. Vigan\`o et al.]{Daniele Vigan\`o$^1$, Diego F.~Torres$^{1,2}$, Kouichi Hirotani$^3$ \& Mart\'in E.  Pessah$^4$\\ 
$^1$Institute of Space Sciences (CSIC--IEEC), Campus UAB, Faculty of Science, Torre C5-parell, E-08193 Barcelona, Spain\\
$^2$Instituci\'o Catalana de Recerca i Estudis Avan\c{c}ats (ICREA) Barcelona, Spain\\
$^3$Academia Sinica, Institute of Astronomy and Astrophysics (ASIAA), PO Box 23-141, Taipei, Taiwan\\
$^4$Niels Bohr International Academy, Niels Bohr Institute, Blegdamsvej 17, DK-2100, Copenhagen \O, Denmark}

\date{}
\maketitle

\label{firstpage}

\begin{abstract}
We consider the fundamental problem of charged particles moving along and around a curved magnetic field line, revising the synchro-curvature radiation formulae introduced by \cite{cheng96}. We provide more compact expressions to evaluate the spectrum emitted by a single particle, identifying the key parameter that controls the transition between the curvature-dominated and the synchrotron-dominated regime. This parameter depends on the local radius of curvature of the magnetic field line, the gyration radius, and the pitch angle. We numerically solve the equations of motion for the emitting particle by considering self-consistently the radiative losses, and provide the radiated spectrum produced by a particle when an electric acceleration is balanced by its radiative losses, as it is  assumed to happen in the outer gaps of pulsar's magnetospheres. 
We compute the average spectrum radiated throughout the particle trajectory finding that the slope of the spectrum before the peak depends on the location and size of the emission region. We show how this effect could then lead to a variety of synchro-curvature spectra.
Our results reinforce the idea that the purely synchrotron or curvature losses are, in general, inadequate to describe the radiative reaction on the particle motion, and the spectrum of emitted photons. Finally, we discuss the applicability of these calculations to different astrophysical scenarios.
\end{abstract}

\begin{keywords}
radiation mechanisms: non-thermal -- gamma-rays: stars -- stars: pulsars: general
\end{keywords}

\section{Introduction}
\label{sec:acceleration}

The motion of charged particles embedded in electric and magnetic fields, $\vec E$ and $\vec B$, is a fundamental problem which is involved in many different astrophysical scenarios. Analytic solutions to the equations of motion can be found in a few special cases: uniform $\vec{E}$ and $\vec{B}$, with $\vec{E}\parallel \vec{B}$ or $\vec{E}\perp \vec{B}$ (see Chap.~12 of \citealt{jackson91}). In a uniform magnetic field, $\vec{B}$, and zero electric field, $\vec{E}=0$, the motion of charged particles is helicoidal, with constant velocity $v_\parallel$ along $\hat{b}$ (the direction of $\vec{B}$), and spiraling around the magnetic field line with cyclotron pulsation, which, in the observer frame, is given by
\begin{equation}
\omega_B = \frac{|q|B}{\Gamma m c}~,
\end{equation}
where $\Gamma\equiv (1 - \beta^2)^{-1/2}$ is the relativistic Lorentz factor, $\beta=v/c$, $v$ is the velocity of the particle, $q$ its charge (multiple of the elementary charge $e$), and $c$ is the speed of light. In the non-relativistic case, $\Gamma \simeq 1$, the cyclotron pulsation is independent on the energy of the particle. Particles with larger perpendicular velocities, $v_\perp$, will spiral around magnetic field lines with a larger {\em Larmor radius} (or {\em gyro-radius}), defined as:
\begin{equation}\label{eq:larmor}
 r_{\rm gyr} = \frac{\Gamma v_\perp m c}{|q|B} =  \frac{\Gamma \beta\sin\alpha \, m c^2}{|q|B}~,
\end{equation}
where the {\it pitch angle} $\alpha$ is defined as the angle between $\vec{B}$ and $\vec{v}$. For an electron, the cyclotron pulsation and the gyro-radius are given by
\begin{eqnarray}
 && \omega_B \simeq 1.76 \left(\frac{B_{12}}{\Gamma}\right) \times 10^{19}~{\rm s}^{-1} = \frac{11.6~B_{12}}{\hbar \Gamma}  ~{\rm keV}~,\\
 && r_{\rm gyr}=\frac{mc^2\beta\Gamma\sin\alpha}{eB} \simeq \left(\frac{\beta\Gamma_7\sin\alpha}{B_6}\right) 1.7 \times 10^4 {\rm cm}~,\label{eq:def_rgyr}
\end{eqnarray}
where, hereafter, we will use the notation $A_x=A/10^x$, with $A$ in cgs units.

Classical electrodynamics ignores the quantum Landau levels. Thus. it works when the energy associated with the perpendicular motion is much larger than the minimum Landau level, or, i.e., when $r_{\rm gyr}$ is much larger than the electron Compton length-scale $\lambda_C\equiv h/m_ec=2.4 \times 10^{-12}$ cm.

In astrophysics, there are many scenarios where particles can be accelerated up to ultra-relativistic velocities and move in a magnetized environment: radiative losses due to the acceleration are very important, especially if the motion of the particle is curved \citep{jackson91}. One can consider two simple kinds of circular motion: sliding along a curved magnetic field line, with radius of curvature $r_c$, and gyration around a straight magnetic field line. For the former, the energy losses by {\it curvature radiation} are (in gaussian units)
\begin{equation}\label{eq:power_curv}
 P_c = \frac{2}{3}\frac{(Ze)^2c \Gamma^4}{r_c^2}~.
\end{equation}
When the gyration is considered, we obtain the {\it synchrotron radiation} power:
\begin{equation}\label{eq:power_sync}
 P_{\rm sync} = \frac{2}{3}\frac{(Ze)^4B^2(\Gamma^2-1)\sin^2\alpha}{m^2c^3}~.
\end{equation}
Synchrotron and curvature radiation are the main candidates to explain the high-energy photons detected in different astrophysical scenarios, such as $\gamma$-ray pulsars. For the latter, much effort has been devoted to develop gap models that account for the observed radiation (e.g., \citealt{sturrock71,ruderman75,arons79,arons83,cheng86a,daugherty96,muslimov03}). However, in general, particles in the gap are electrically accelerated along the magnetic field lines, but, at the same time, they spiral around them. In this sense, one has to worry about a careful evaluation of the particle dynamics. This led \cite{cheng96} to first propose a fully consistent formulation of the synchro-curvature radiation, later studied also by \cite{zhang00}. \cite{kelner12} and \cite{prosekin13} accurately calculate the dynamics and synchro-curvature radiation (labeled as 'magnetic bremmstrahlung') produced by particles moving in a dipolar magnetic field, and discuss the black hole magnetospheres and the pulsar gaps as applications of the formulae. \cite{zhang11} evaluate the self-synchro-curvature Compton radiation, evaluating its impact in different astrophysical scenarios, like AGNs and GRBs. However, in general, the synchro-curvature radiation is overlooked, in favour of a purely synchrotron radiation.

In this work we aim at revising these calculations, finding an equivalent, more compact way to calculate the radiative losses. While the formulation is formally equivalent to the original work by \cite{cheng96}, our notation allows the identification of a key parameter, related to the radius of curvature of the line, the magnetic field and the pitch angle. Such parameter regulates the transition from the  curvature-dominated to the synchrotron-dominated regimes, and allows us to assess whether a purely synchrotron or curvature formulae approximates well the radiative losses.

In \S\ref{sec:synchrocurvature} we revise the formulae for the synchro-curvature losses, writing them in a much more compact form and identifying the above-mentioned parameter. In \S\ref{sec:motion} we solve the equations of motion considering radiative losses in a self-consistent way, and in \S\ref{sec:spectra} we show the spectra emitted by the particles along their trajectories. In \S\ref{sec:discussion} we comment on our results and discuss the possible applications, beyond the obvious case of the outer gap for pulsars.

\section{Compact formulae for synchro-curvature losses}\label{sec:synchrocurvature}

\cite{cheng96} consider the {\em synchro-curvature radiation}, i.e., the radiation emitted by a particle spiraling around a circular magnetic field line with radius of curvature $r_c$ and magnetic field intensity $B$. Their calculations hold as long as $r_c\gg r_{\rm gyr}$, and $r_c$ changes on length-scales much larger than $r_{\rm gyr}$. These assumptions are safe in the outer pulsar magnetosphere, since $r_c$ is of the order of the light cylinder ($r_c\sim cP/2\pi \sim 10^7-10^9$ cm).

\cite{cheng96} obtain that particles emit photons with a characteristic energy
\begin{equation}\label{eq:characteristic_energy}
 E_c(\Gamma,r_c,r_{\rm gyr},\alpha) = \frac{3}{2}\hbar cQ_2\Gamma^3~.
\end{equation}
Here, the factor $Q_2$ is defined by \cite{cheng96}:
\begin{equation}\label{eq:q2_simple}
 Q_2^2 = \frac{\cos^4\alpha}{r_c^2}\left[1 + 3\xi  + \xi^2 + \frac{r_{\rm gyr}}{r_c}\right]~,
\end{equation}
where we have defined the synchro-curvature parameter
\begin{equation}\label{eq:def_xi}
   \xi = \frac{r_c}{r_{\rm gyr}}\frac{\sin^2\alpha}{\cos^2\alpha} \simeq 5.9\times 10^3~ \frac{r_{c,8} B_6 \sin\alpha}{\Gamma_7 \cos^2\alpha} ~.
\end{equation}
Since $r_{\rm gyr}\ll r_c$ by assumption, the last term in the right-hand side of eq.~(\ref{eq:q2_simple}) is always negligible compared with the first three. Therefore, in order to compare the curvature and synchrotron contributions, the fundamental parameter to consider is $\xi$. Note that the ratio between the purely curvature and synchrotron powers, eqs.~(\ref{eq:power_curv}) and (\ref{eq:power_sync}), is (for $\Gamma \gg 1$)
\begin{equation}
 \frac{P_c}{P_{\rm sync}} = \xi^2 \frac{\cos^2\alpha}{\sin^2\alpha} = \frac{\xi^2}{\tan^2\alpha}~.
\end{equation}
The power radiated by one particle per unit energy is \citep{cheng96}:

\begin{figure}
\centering
\includegraphics[width=.49\textwidth]{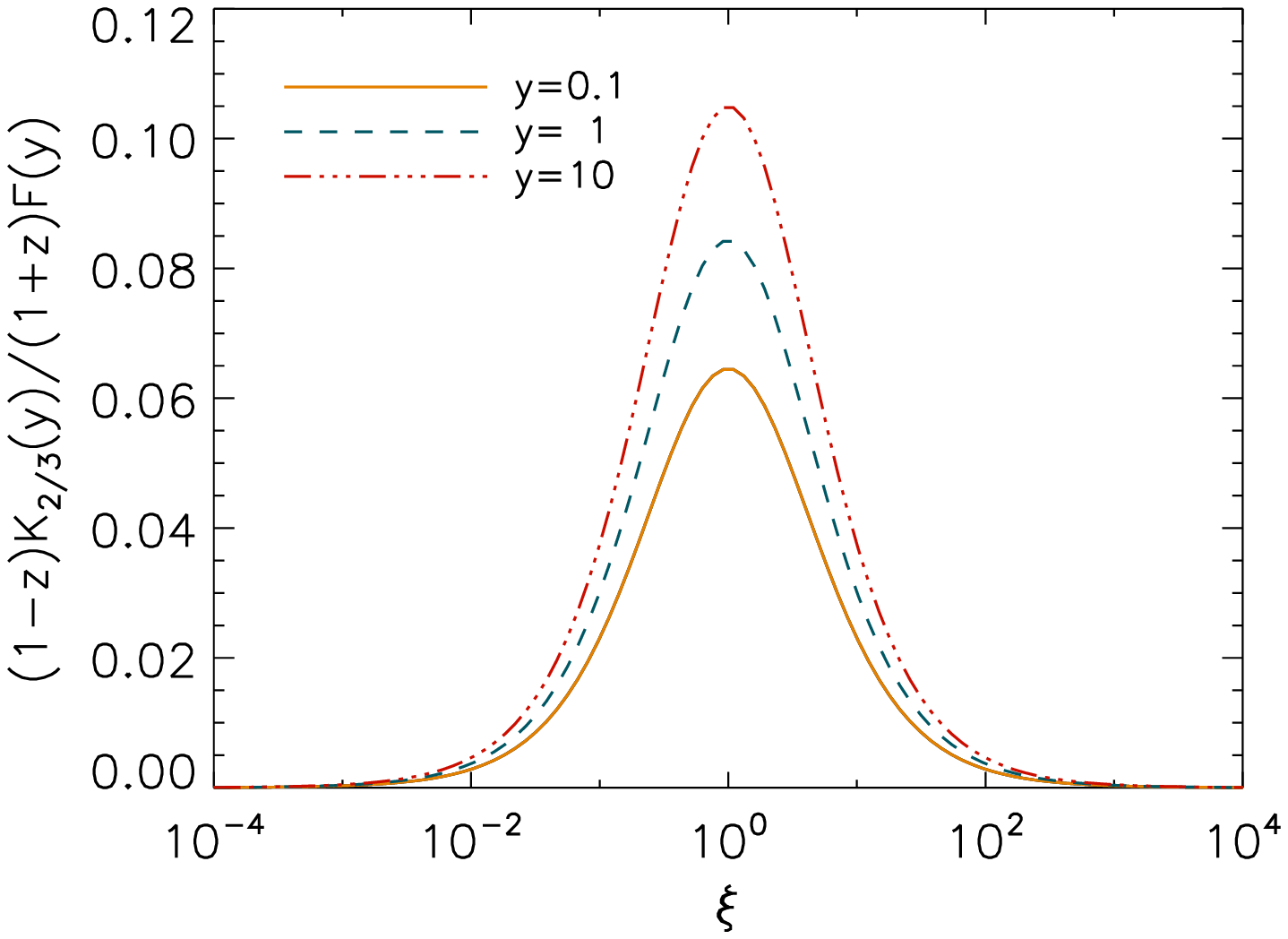}\\
\includegraphics[width=.49\textwidth]{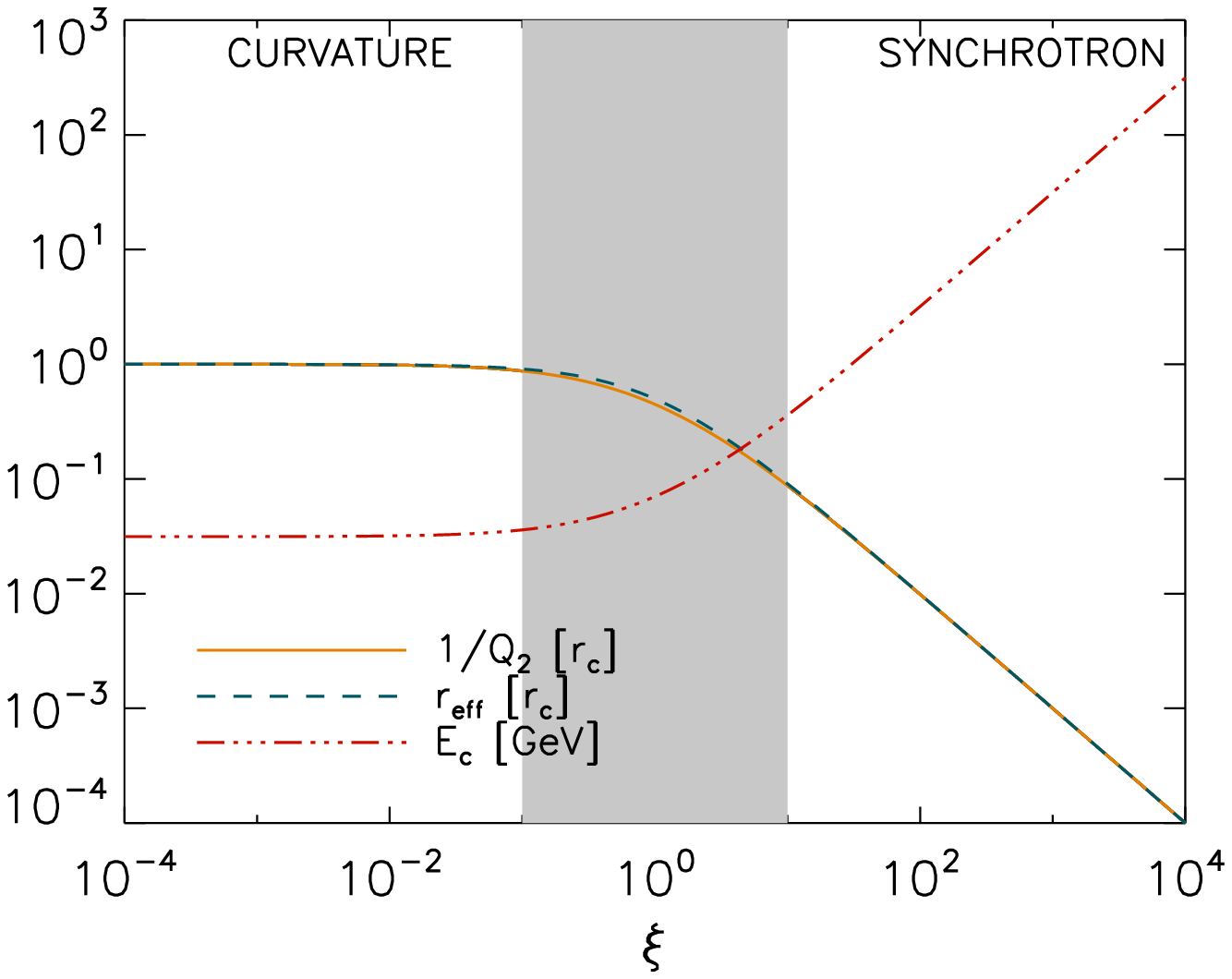}\\
\includegraphics[width=.49\textwidth]{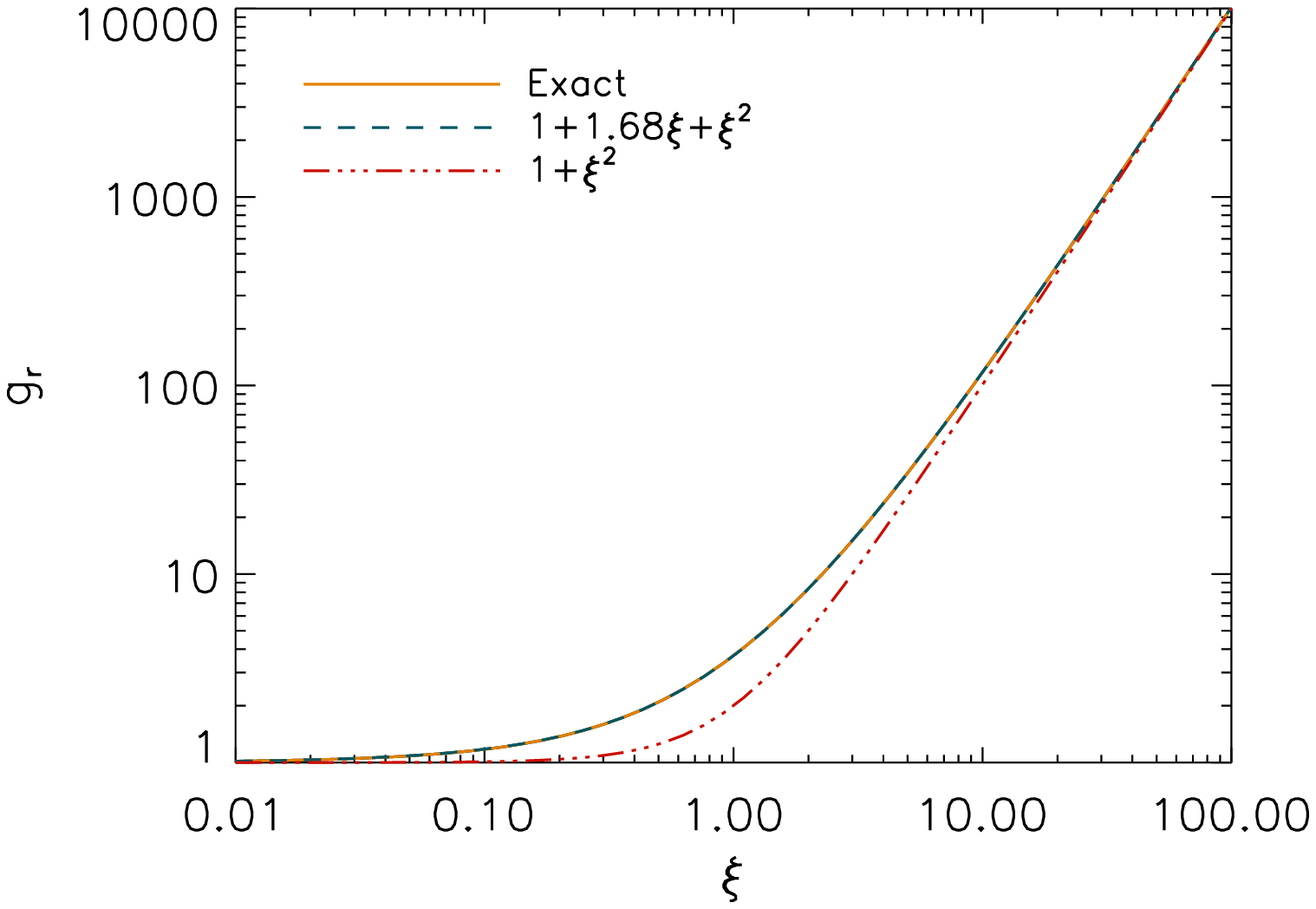}
\caption{Synchro-curvature parameters as a function of $\xi$, for $\Gamma=10^7$, $r_{\rm gyr}=1$ cm and $r_c=10^8$ cm. Top: ratio between the terms of the r.h.s. of eq.~(\ref{eq:sed_synchrocurv}), $(1 - z) K_{2/3}(y)/(1 + z) F(y)$, for three different values of $y=E/E_c$, with $\Gamma=10^7$, $r_{\rm gyr}=1$ cm and $r_c=10^8$ cm. Middle: values of $Q_2^{-1}$, $r_{\rm eff}$ and $E_c$. We indicate the regimes of curvature and synchrotron domination, and the grey band indicates the region of transition. Bottom: the factor $g_r=P_{sc}/P_c$, compared with two analytical fits: $1+1.68\xi + \xi^2$ and $1 + \xi^2$.}
 \label{fig:q2}
\end{figure}

\begin{equation}\label{eq:sed_synchrocurv}
 \frac{dP_{sc}}{dE} = \frac{\sqrt{3} (Ze)^2 \Gamma y}{4\pi \hbar r_{\rm eff} }[(1 + z) F(y) - (1 - z) K_{2/3}(y) ]~,
\end{equation}
where 
\begin{eqnarray}
 && y(E,\Gamma,r_c,r_{\rm gyr},\alpha) \equiv \frac{E}{E_c}~,\\
 && z = (Q_2 r_{\rm eff})^{-2} ~,\\
 && F(y) = \int_y^\infty K_{5/3}(y') dy'~,
\end{eqnarray}
and $K_n$ are the modified Bessel functions of the second kind of index $n$. We have defined the effective radius as:
\begin{equation}\label{eq:def_reff}
r_{\rm eff} = \frac{r_c}{\cos^2\alpha}\left(1 + \xi+ \frac{r_{\rm gyr}}{r_c}  \right)^{-1}~.
\end{equation}
In the top panel of Fig.~\ref{fig:q2}, we show the ratio between the two terms appearing on the right-hand side of eq.~(\ref{eq:sed_synchrocurv}), for different values of $y$. The second term $\propto K_{2/3}$ is negligible, unless $\xi \sim 1$ and $y\gtrsim 1$, for which its value represents a $\sim 10\%$ of the first term. For larger values of $y$, this weight increases, but the fraction of photons with such large energies is small.

The total power radiated by one particle can be found by integrating eq.~(\ref{eq:sed_synchrocurv}) in energy \citep{cheng96}:
\begin{equation}\label{eq:power_synchrocurv}
 P_{sc} = \frac{2(Ze)^2 \Gamma^4 c}{3 r_c^2} g_r~,
\end{equation}
where we have introduced the synchro-curvature correction factor
\begin{equation}\label{eq:def_gr}
 g_r =  \frac{r_c^2}{r_{\rm eff}^2}\frac{[1 + 7(r_{\rm eff}Q_2)^{-2}]}{8 (Q_2r_{\rm eff})^{-1}}~,
\end{equation}
which takes into account the deviation from the curvature power, eq.~(\ref{eq:power_curv}).

Note that our eqs.~(\ref{eq:q2_simple}), (\ref{eq:def_reff}), (\ref{eq:sed_synchrocurv}) and (\ref{eq:power_synchrocurv}) are equivalent to eqs.~(3.6), (3.27), (3.32), and (3.35), respectively, of \cite{cheng96}. However, our formulation is more compact and allows a quick evaluation of whether we are in either the purely curvature or synchrotron limits.\footnote{The equivalence can be proven by substituting the explicit expressions $g_r$ and $\xi$ in our formalism. In particular, \cite{cheng96} define $r_{\rm eff} = c^2/[(r_{\rm gyr}+r_c) \Omega_0^2  + r_{\rm gyr}\omega_B^2]$, where they introduce the effective frequency $\Omega_0 = c \cos\alpha/r_c$. To simplify their eq.~(3.35), we have evaluated the Gamma functions: $\Gamma(7/3)\Gamma(2/3)=8\pi/3^{5/2}$ and $\Gamma(4/3)/\Gamma(7/3)=3/4$ since $\Gamma(n)\equiv n!$ by definition. In the same equation, there is a misprint: $r_c$ should be substituted by $r_{\rm eff}$ ($r_c^\star$ in their notation).}

If $\xi \ll 1$, then $g_r=1$ and the curvature radiation dominates, and eq.~(\ref{eq:power_curv}) is recovered, since the following limits hold:
\begin{eqnarray}
&& \lim_{\xi \ll 1} Q_2 = \frac{1}{r_c}~, \label{eq:q2_lim_curv}\\
&& \lim_{\xi \ll 1} r_{\rm eff} = r_c~, \\
&& \lim_{\xi \ll 1} g_r = 1~, \\
&& \lim_{\xi \ll 1} E_c = E_{\rm curv}(\Gamma,r_c) = \frac{3}{2} \frac{ \hbar c}{r_c} \Gamma^3 ~,\\
&& \lim_{\xi \ll 1} P_{sc} = P_c~. \label{eq:power_lim_curv}
\end{eqnarray}
Note that such limits correspond to very small values of $\alpha \ll 10^{-4}\Gamma_7/B_6r_{c,8}$, i.e., for a motion mainly consisting in a sliding of particles along the magnetic field line.

On the other hand, if $\xi \gg 1$, then synchrotron losses dominate, and eq.~(\ref{eq:power_synchrocurv}) reduces to eq.~(\ref{eq:power_sync}):
\begin{eqnarray}\label{eq:characteristic_energy_sync}
&& \lim_{\xi \gg 1} Q_2 = \frac{\xi}{r_c} = \frac{\sin^2\alpha}{r_{\rm gyr}} = \frac{eB  \sin\alpha}{\Gamma m c^2}~, \\
&& \lim_{\xi \gg 1} r_{\rm eff} = \frac{r_{\rm gyr}}{\sin^2\alpha} = \frac{mc^2\Gamma}{eB\sin\alpha}~, \\
&& \lim_{\xi \gg 1} g_r =  r_c^2 \left(\frac{eB\sin\alpha}{mc^2\Gamma}\right)^2~, \\
&& \lim_{\xi \gg 1} E_c = E_{\rm sync}(\Gamma,\alpha,B) = \frac{3}{2} \Gamma^2 \frac{\hbar eB \sin\alpha}{mc} ~,\\
&& \lim_{\xi \gg 1} P_{sc} = P_{\rm sync}~.
\end{eqnarray}
In the middle panel of Fig.~\ref{fig:q2} we show the calculated values of $(Q_2)^{-1}$, $r_{\rm eff}$ and $E_c$ as a function of $\xi$, for $\Gamma=10^7$, $r_{\rm gyr}=1$ cm (implying $\sin\alpha/B_6\simeq 6\times 10^{-5}$, eq.~\ref{eq:def_rgyr}), and $r_c=10^8$ cm. The value of $z\equiv (Q_2 r_{\rm eff})^{-2}$ is always $z\leq 1$, and in both limits $\xi \ll 1$ or $\xi \gg 1$, $z\rightarrow 1$. For $\xi =1$, if $\cos\alpha\rightarrow 0$ and $r_{\rm gyr}\ll r_c$, then $z=0.8$. The region of transition between the two limits spans roughly two orders of magnitude in $\xi$ (grey band).

In the bottom panel we show the factor $g_r=P_{sc}/P_c$ as a function of $\xi$, for the same sets of parameters. We overplot two analytical approximations: $1+\xi^2$ and $1 + 1.68\xi + \xi^2$. The former represents the simple approximation $P_{sc}=P_c + P_{\rm sync}$, and it is accurate only within a factor of a few if $\xi\sim 1$. The second form is the best parabolic fit, which reproduces the exact values within $\lesssim 0.3\%$. Note that this analytical approximation would be slightly worse for non-negligible values of $\sin\alpha$. On the other hand, all the results shown in Fig.~\ref{fig:q2} are valid for any value of $\Gamma$ and $r_{\rm gyr}$, as far as $r_{\rm gyr}\ll r_c$, which is an assumption in the work of \citealt{cheng96} to derive the synchro-curvature formulae.

\begin{table} 
\caption{Assumed parameters for the calculation of particle dynamics (see Fig.~\ref{fig:tra2d}).}
\begin{center}
\begin{tabular}{ccccc} 
\hline 
\hline
Model & $E_\parallel$ & $B$ & $r_c$ & $\Gamma_{\rm in}$ \\
 & [V/m] & [G] & [cm] & \\
\hline
A & $10^7$ & $10^6$ & $10^8$ & $10^3$ \\
B & $10^8$ & $10^6$ & $10^8$ & $10^3$ \\
C & $10^7$ & $10^8$ & $10^8$ & $10^3$ \\
D & $10^7$ & $10^6$ & $10^8$ & $10^2$ \\
E & $10^7$ & $10^6$ & $10^7$ & $10^3$ \\
\hline 
\end{tabular} 
\label{tab:models} 
\end{center}
\end{table} 

Last, we remark that the expression (\ref{eq:power_synchrocurv}) represents the emitted power integrated over the solid angle. However, we remind that the radiation is beamed within a cone characterized by a semi-aperture $\theta_c$, which is approximately given by (see, e.g., Chapter 14 of \citealt{jackson91})

\begin{eqnarray}
 && \theta_c = \frac{1}{\Gamma}\left( \frac{2E_c}{E} \right)^{1/3} {\qquad \rm for } E\ll E_c \label{eq:beam1}\\
 && \theta_c = \frac{1}{\Gamma}\left( \frac{2E_c}{3E} \right)^{1/2} {\qquad \rm for } E\gg E_c  ~, \label{eq:beam2}
\end{eqnarray}
where $E_c \propto \Gamma^3$, eq.~(\ref{eq:characteristic_energy}). This has strong implications for the dependence of spectra on the viewing angle. For a given $\Gamma$, photons with higher energies will be emitted within a smaller beam. At the peak of the spectrum, $E \sim E_c$, Eqs.~(\ref{eq:beam1}) and (\ref{eq:beam2}) tell us that the larger $\Gamma$, the narrower the beam. As a consequence, an off-axis observer will detect a softer radiation, compared with the on-axis observer, for two reasons: the larger ratio between soft and hard photon fluxes; more radiation coming from low-$\Gamma$ particles, which peaks at lower $E_c$. On top of that, the geometry of the magnetic field lines and the viewing angles will play a dominant role, selecting those parts of the particle trajectories where the detected radiation is originated.

\section{Particles motion}\label{sec:motion}

Let us now suppose that a particle is subject to an electric field $E_\parallel$, directed along $\hat{b}$, i.e., tangential to the magnetic field line. As we deal with relativistic particles only, their velocity remains constant ($v=c$) so that the electric force in only used to increase the energy, with no effect on the velocity. Therefore, there is almost no parallel acceleration $dv/dt$, hence no associated reaction force. The radiation emitted in the direction of motion, $\hat{p}$, implies a reaction force $\vec{F}_{\rm rad} = (P_{sc}/v)~ \hat{p}$. This is a good approximation in the pulsar gaps, where these two forces are much stronger than the gravitational and hydrodynamical ones. The equation for the particle relativistic momentum, $\vec{p} \equiv \sqrt{\Gamma^2 -1}mc~ \hat{p} = \Gamma mv ~\hat{p}$, is related to the synchro-curvature power, eq.~(\ref{eq:power_synchrocurv}), by:

\begin{figure}
\centering
\includegraphics[width=.5\textwidth]{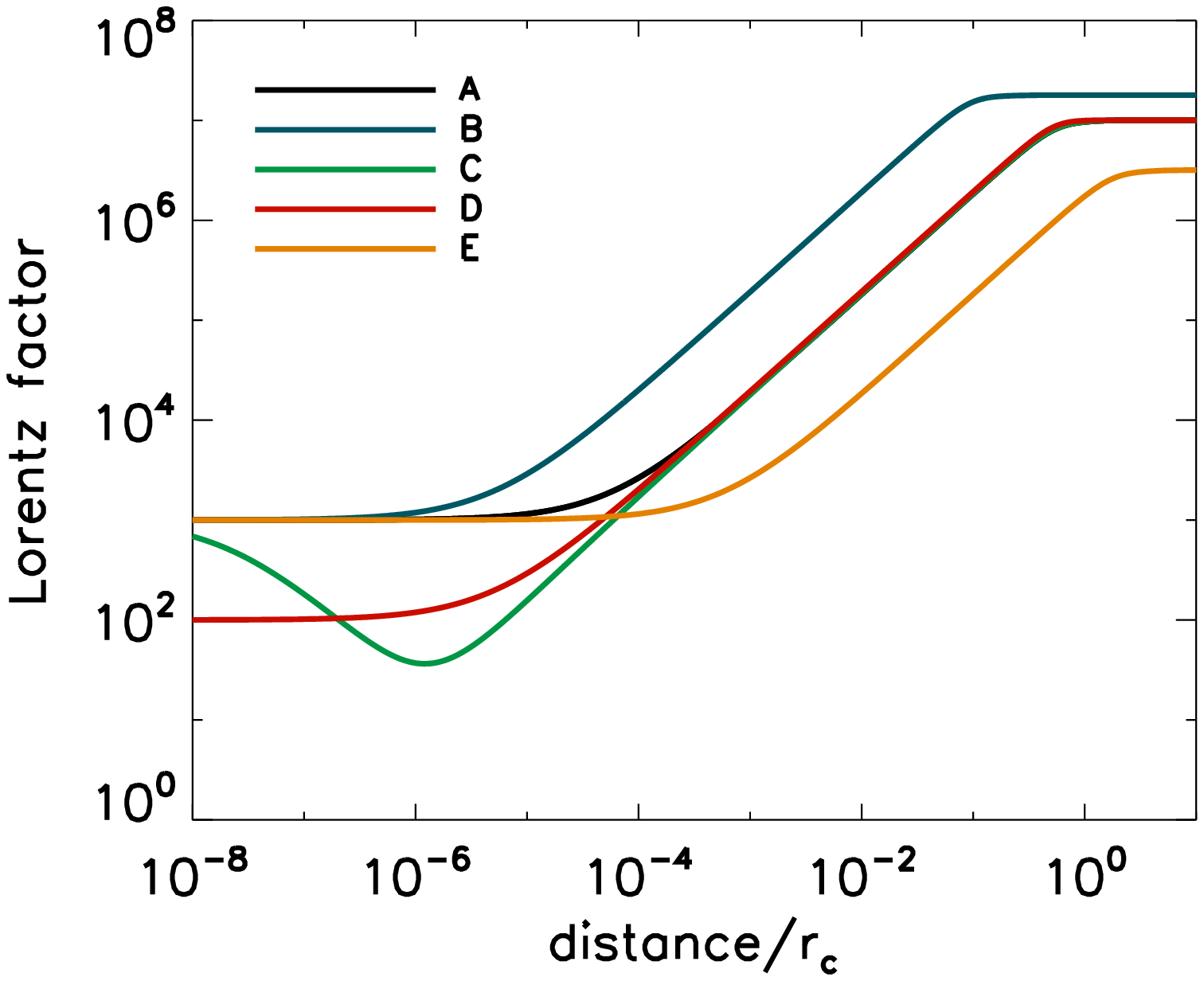}\\
\includegraphics[width=.5\textwidth]{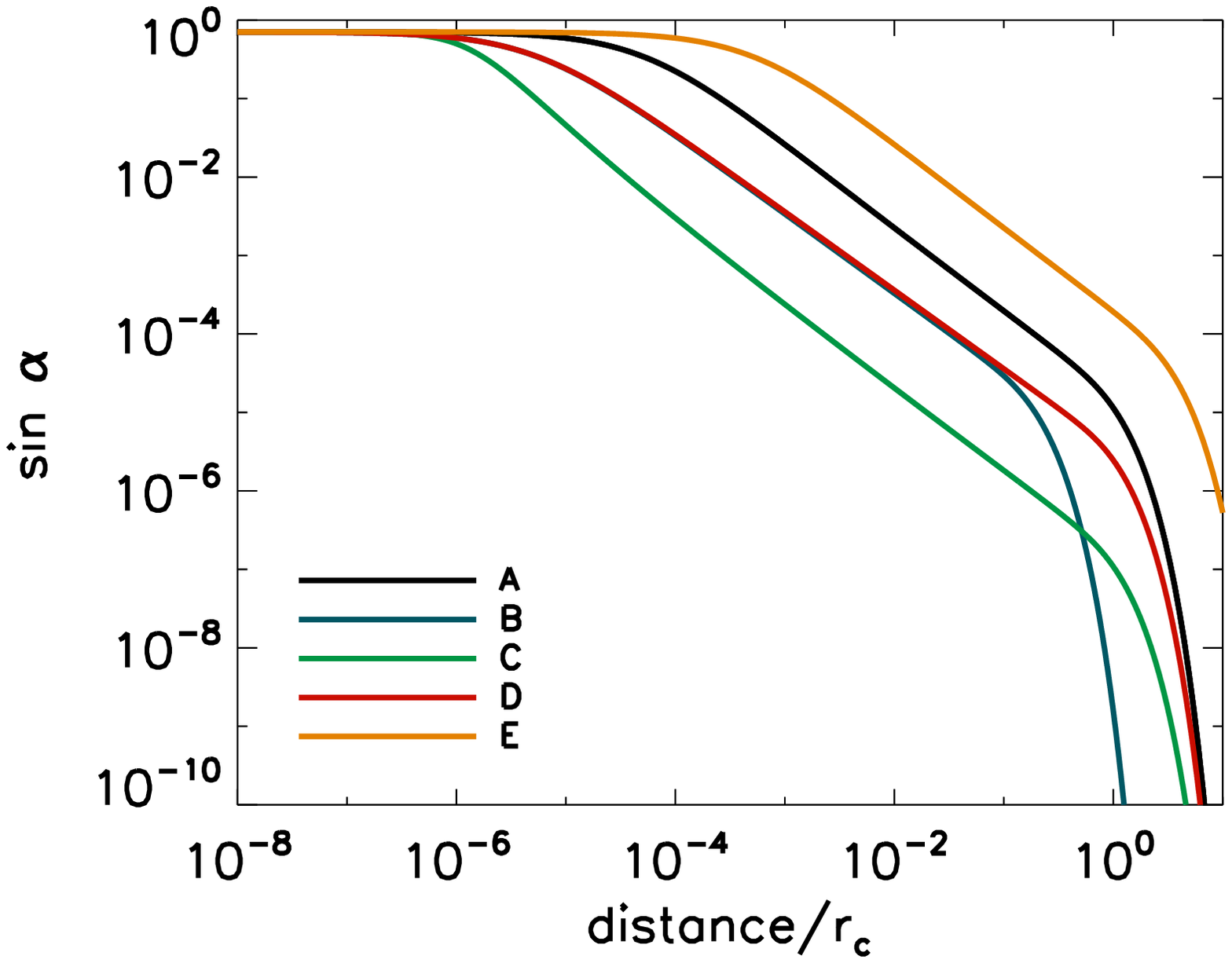}\\
\includegraphics[width=.5\textwidth]{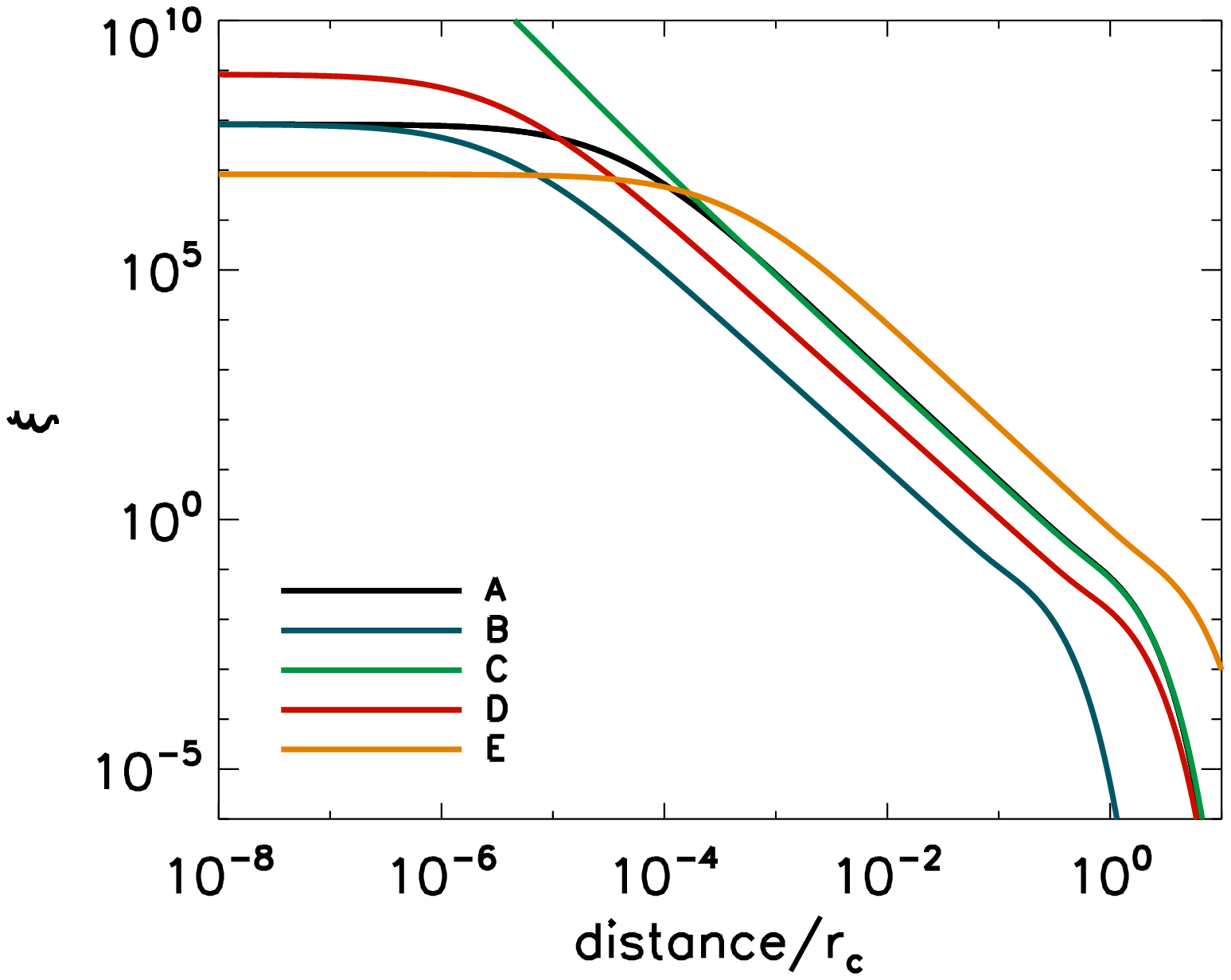}
\caption{Evolution of $\Gamma$ (top), $\sin\alpha$ (middle) and $\xi$ (bottom), for a charged particle with $qE_\parallel > 0$, as a function of the distance along a magnetic field line (normalized by $r_c$, and where the origin is the place of particle creation), for the models of Table~\ref{tab:models}. We assume an initial pitch angle $\alpha_{\rm in}=\pi/4$.}
 \label{fig:tra2d}
\end{figure}

\begin{figure}
\centering
\includegraphics[width=.5\textwidth]{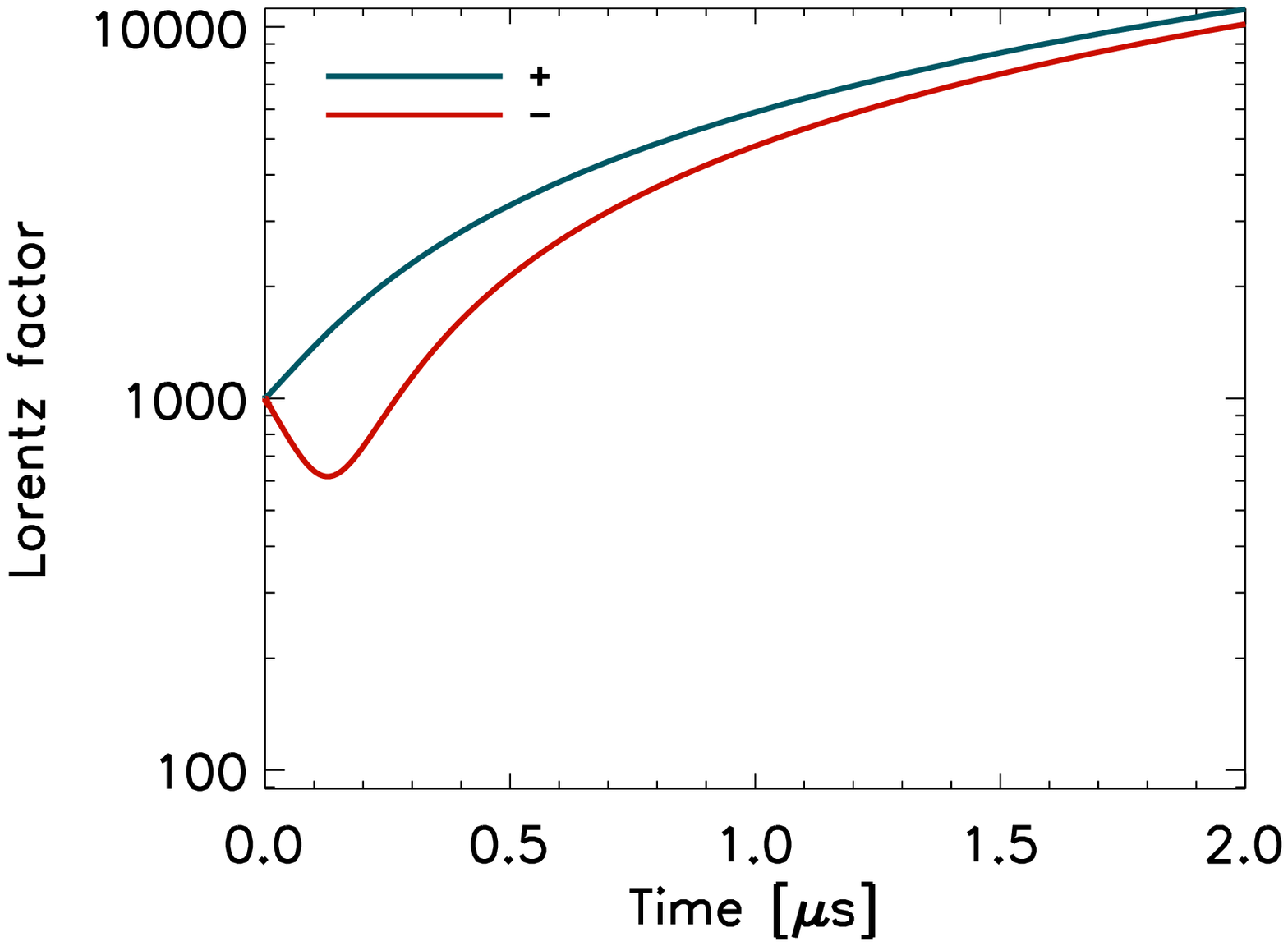}\\
\includegraphics[width=.5\textwidth]{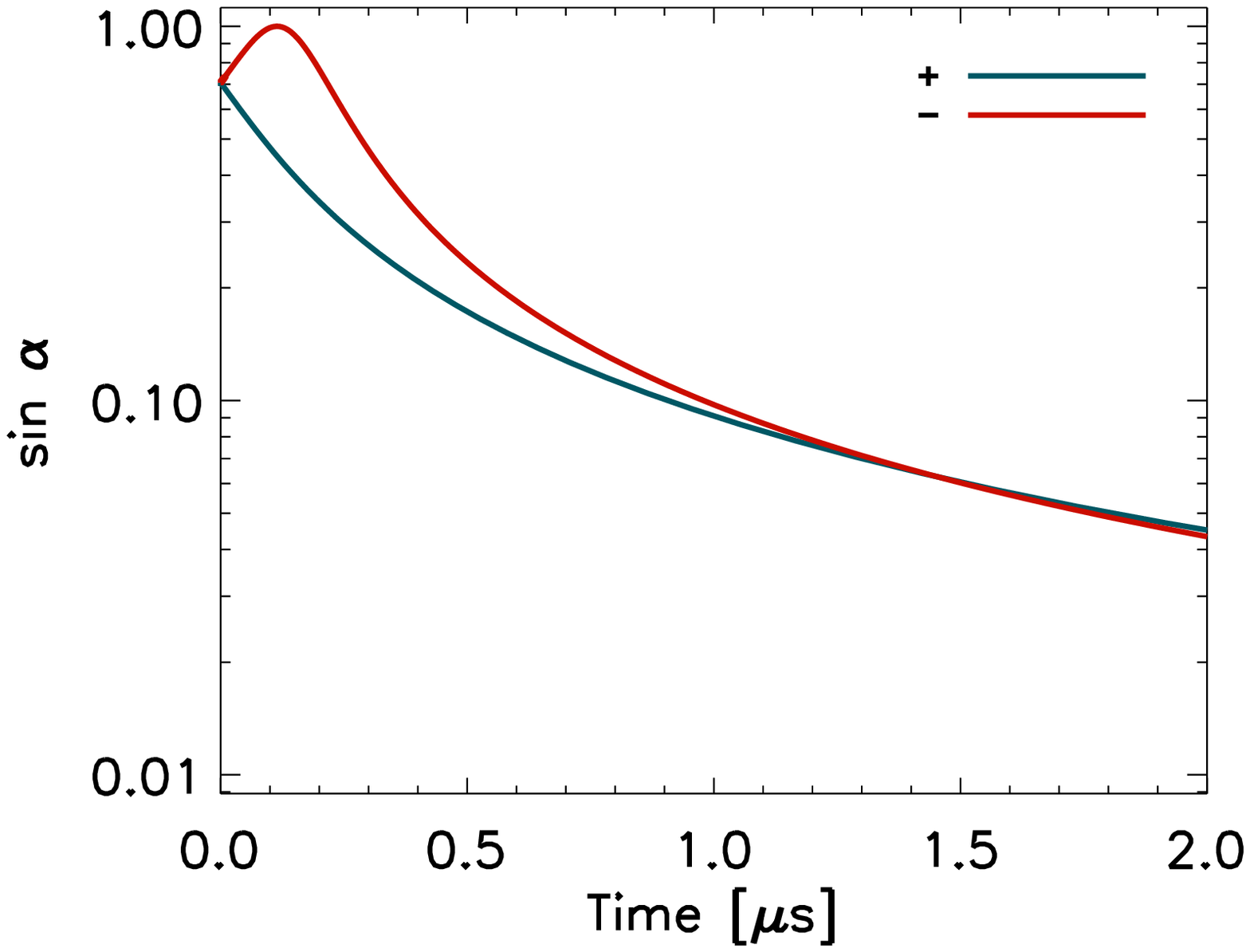}\\
\includegraphics[width=.5\textwidth]{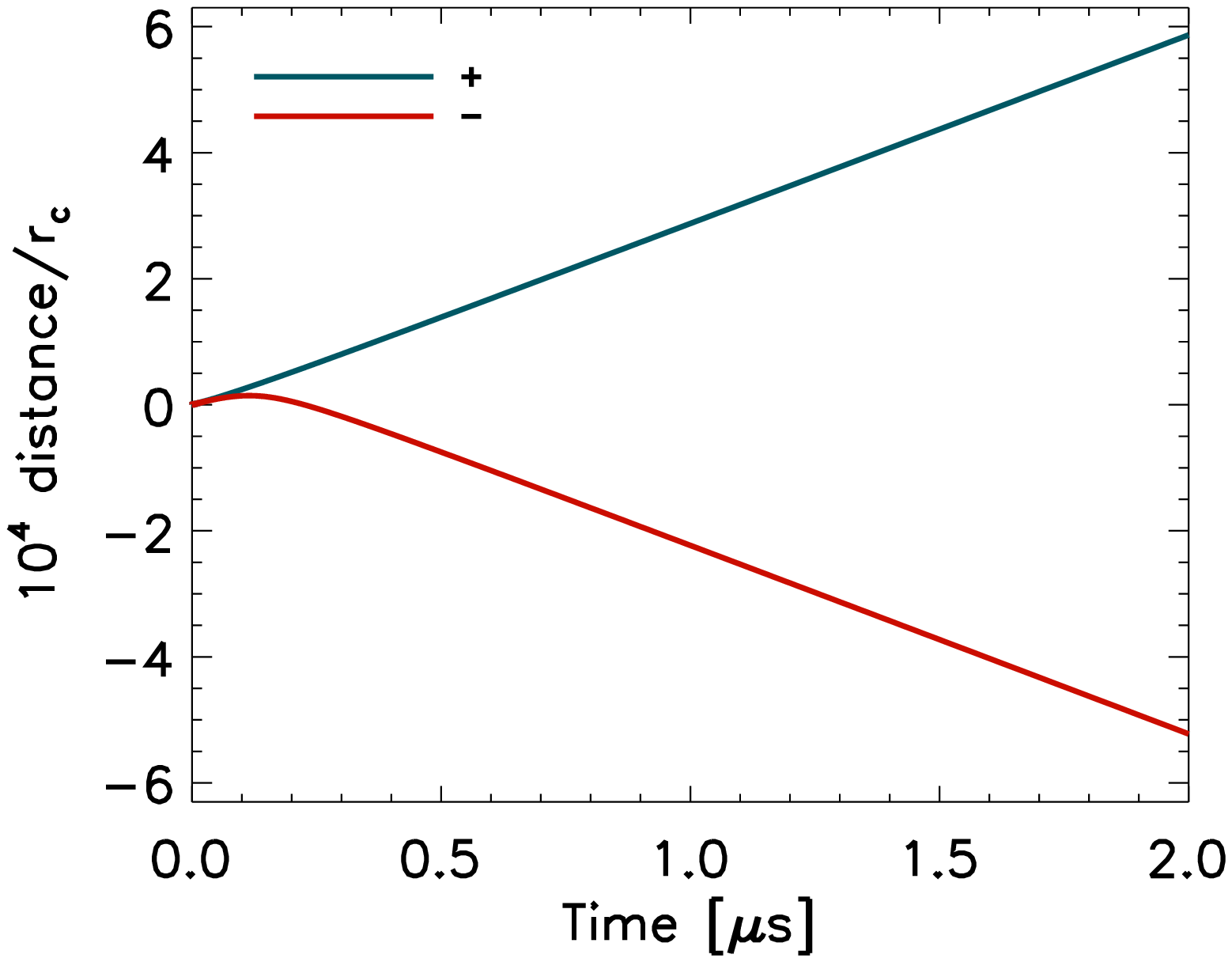}
\caption{Early evolution of $\Gamma$ (top), $\sin\alpha$ (middle) and position (bottom), for a positive (blue) or negative (red) value of $qE_\parallel$, as a function of the time since particle creation, for the model A of Table~\ref{tab:models}. We assume an initial pitch angle $\alpha_{\rm in}=\pi/4$.}
 \label{fig:reversal}
\end{figure}

\begin{equation}
 \frac{d\vec{p}}{dt} = ZeE_\parallel \hat{b} - \frac{P_{sc}}{v}~\hat{p}~.
\end{equation}
Decomposing the motion into parallel ($p_\parallel=p\cos\alpha$) and perpendicular components ($p_\perp = p\sin\alpha$), we obtain:\footnote{Eqs.~(\ref{eq:motion_perp}) and (\ref{eq:motion_par}) are a generalization of eqs.~(56) and (57) of \cite{hirotani99a}, where they consider $P_{sc}=P_{\rm sync}$.}
\begin{eqnarray}
 && \frac{\de(p\sin\alpha)}{\de t} = - \frac{P_{sc}\sin\alpha}{v}~, \label{eq:motion_perp} \\
 && \frac{\de(p\cos\alpha)}{\de t} = ZeE_\parallel - \frac{P_{sc}\cos\alpha}{v}~. \label{eq:motion_par}
\end{eqnarray}
Let us consider electrons and positrons that are created by pair production (like in the outer gap), so that they have an initial energy $E_{\rm in}=\Gamma_{\rm in}mc^2$ that depends on the $\gamma$-ray photon energy:

\begin{equation}\label{eq:gammain}
  \Gamma_{\rm in} \sim \frac{E_\gamma}{2m_ec^2} \simeq 10^3 E_\gamma{\rm [GeV]}~.
\end{equation}
The pitch angle at birth, $\alpha_{\rm in}$, depends, in general, on the trajectories and energies of the progenitor photons, their mean free path and the local curvature of magnetic field lines. In general, one can expect $\alpha_{\rm in}\sim $ O(1) (see, e.g., \citealt{takata06}), which implies $\xi_{\rm in}\gg 1$. In Table~\ref{tab:models} we show five sets of values for the model parameters $E_\parallel, B, r_c, \Gamma_{\rm in}$. For all of them, we fix the initial pitch angle to $\alpha_{\rm in}=\pi/4$. 

In Fig.~\ref{fig:tra2d} we show the results of the numerical integration of eqs.~(\ref{eq:motion_perp}) and (\ref{eq:motion_par}), for a particle with $qE_\parallel > 0$, and the initial parallel velocity directed as the electric force, and initial pitch angle $\alpha_{\rm in}=\pi/4$. We show the evolution of $\Gamma$ (top panel), $\sin\alpha$ (middle), and $\xi$ (bottom), as a function of covered distance (normalized by $r_c$). The dynamical evolution can be divided in three phases: the short synchrotron-dominated regime, the linear growth of $\Gamma$, and the saturation regime.


In Fig.~\ref{fig:reversal} we show what happens in the initial (short) phase, for particles of different sign, using model A as an example. The chosen values of $E_\parallel$ (typical of the outer gap models) are large enough to quickly separate pairs of particle of different sign, and accelerate the electron and the positron towards opposite directions of the magnetic field line (see also \citealt{hirotani99a}). The typical length-scale for charge separation is 
\begin{equation}
 x_{\rm sep} \sim c\frac{p}{eE_\parallel} \sim c\frac{\Gamma_{\rm in} cm_e}{eE_\parallel} = 5\times 10^3 \left(\frac{\Gamma_{\rm in,3}}{E_7}\right) {\rm~cm}~,
\end{equation}
where $E_7=E_\parallel/(10^7$ V/m). Since $\xi_{\rm in}\gg 1$, soon after the pair creation, the losses are dominated by an efficient, purely synchrotron radiation. If $qE_\parallel < 0$, then the particle reverts its motion, which means that the value of $\sin\alpha$ rises to 1, to decrease later.

During this phase, if $\xi$ is large enough (e.g., due to strong $B$, as in model C of Fig.~\ref{fig:tra2d}), then $\Gamma$ temporarily diminishes even if $qE_\parallel > 0$. The timescale of synchrotron losses can be estimated from the ratio between the total energy of the particle and the synchrotron power:

\begin{equation}
 t_{\rm sync} \sim \frac{3m_e^3c^5}{2e^4B^2\sin^2\alpha_{\rm in}\Gamma_{\rm in}} \sim \frac{5\times 10^{-7}}{B_6^{2}\Gamma_{\rm in, 3}\sin^2\alpha_{\rm in}}~{\rm s}~.
\end{equation}
In the second phase, for $x\gtrsim x_{\rm sep}$, the evolution of $\Gamma$ and $\xi$ shown in Fig.~\ref{fig:tra2d} is driven by the kinetic energy provided by the very large $E_\parallel$: the trajectories are specular (if $r_c,E_\parallel,B$ are constant), and what follows holds  independently from the sign of the charge. As $\Gamma$ increases linearly under the electric acceleration, $\sin\alpha$ and $\xi$ decrease linearly and quadratically, respectively. When the value of $\xi$ goes below 1, i.e., $r_{\rm eff} \rightarrow r_c$ and $g_r \rightarrow 1$, the losses are dominated by the curvature radiation and particles are basically directed along $\vec{B}$.

The linear growth of $\Gamma$ halts when the synchro-curvature radiation reaction is effective enough to balance the power given by the electric acceleration. At the same time, the perpendicular motion has been largely dissipated, so that $\sin\alpha\ll 1$. In this saturation regime, eq.~(\ref{eq:motion_par}) (and use also of eq.~(\ref{eq:power_curv})) predicts that the asymptotic value of $\Gamma$ is given by

\begin{equation}\label{eq:gamma_sat}
  \Gamma_{\rm st} = \left( \frac{3}{2}\frac{E_{\parallel, \rm st} r_c^2}{e} \right)^{1/4} \simeq 1.01\times 10^7 (E_7 r_{c,8}^2)^{1/4}~.
\end{equation}
The larger $E_{\parallel, \rm st}$ and/or the larger $r_c^2$ (less efficient radiative losses), the larger $\Gamma_{\rm st}$. Once $\Gamma \rightarrow \Gamma_{\rm st}$ and $\cos\alpha \rightarrow 1$, eq.~(\ref{eq:motion_par}) shows that $p\cos\alpha \sim p$ approaches a constant value, and, consequently, $P_{sc}/v \rightarrow eE_\parallel$. Substituting this in eq.~(\ref{eq:motion_perp}), this implies that the pitch angle decays exponentially, as seen in the middle panel:

\begin{equation}
  \sin\alpha \sim \exp{(-eE_\parallel t/\Gamma_{\rm st} m_e c)}~.
\end{equation}
As a consequence, particles will lie on the fundamental Landau level, and $\xi \rightarrow 0$ (curvature radiation limit).\footnote{This contrasts with the choice of $B,P$-dependent, finite, steady-state value of $\sin\alpha$ by \cite{zhang97} when studying outer gap applications. See the Appendix.} 

The timescale needed to reach the saturated value is

\begin{equation}
 t_{\rm st} \sim \frac{\Gamma_{\rm st}cm_e}{eE_{\parallel, \rm st}} = 1.7\left(\frac{\Gamma_{\rm st,7}}{E_7}\right) {\rm~ms} \simeq 1.7~ \frac{r_{c,8}^{1/2}}{E_7^{3/4}}~ {\rm ms}~,
\end{equation}
where we are considering ultra-relativistic motion, $v_\parallel \sim c$. This corresponds to a distance $x_{\rm st} \simeq ct_{\rm st}$

\begin{equation}
 x_{\rm st} \sim 5.2\times 10^7 \left(\frac{\Gamma_{\rm st,7}}{E_7}\right) {\rm~cm} \sim 5.2 \times 10^7 \frac{r_{c,8}^{1/2}}{g_r^{1/4}E_7^{3/4}} ~{\rm cm~}~.
\end{equation}
If $E_\parallel$ is strong ($\sim 10^8$ V/m, model B), then the distance $x_{\rm st}$ needed by the electric acceleration to bring $\Gamma$ to the steady-state value is only a fraction of $r_c$. However, if $E_\parallel\sim 10^7$ V/m, then $x_{\rm st}\sim r_c$.

Note that the evolution of $\Gamma$ for the models A, C and D converge, soon after the synchrotron-dominated phase. Model B reaches larger $\Gamma$ due to the stronger $E_\parallel$, while the high radiative efficiency of model E (low $r_c$) provides a relatively low saturated value $\Gamma_{\rm st}$. The timescales which separate the three phases are in agreement with the analytical estimates above.

%
%
%

\begin{figure*}
\centering
\includegraphics[width=.52\textwidth]{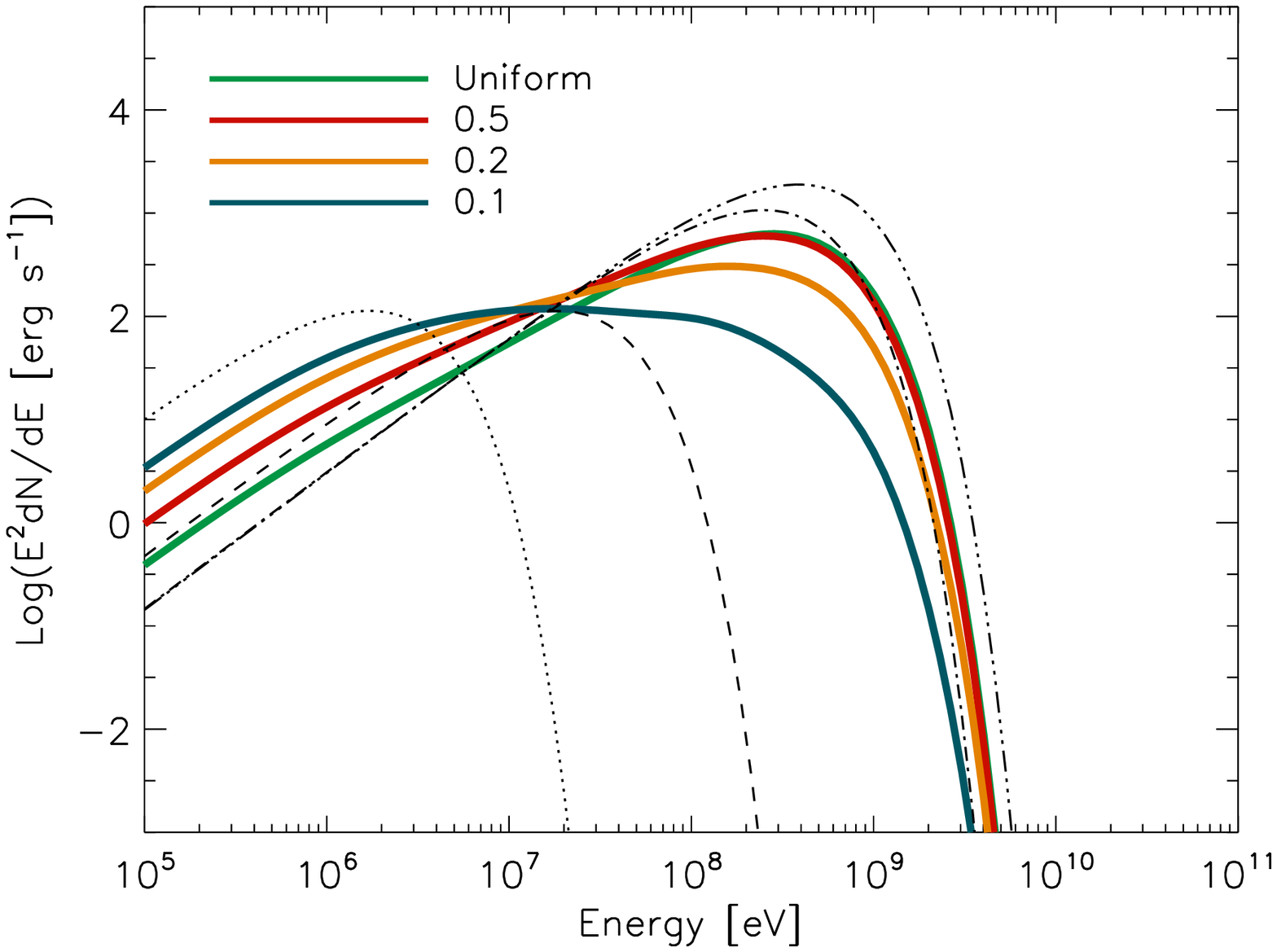}
\hspace{-1cm}
\includegraphics[width=.52\textwidth]{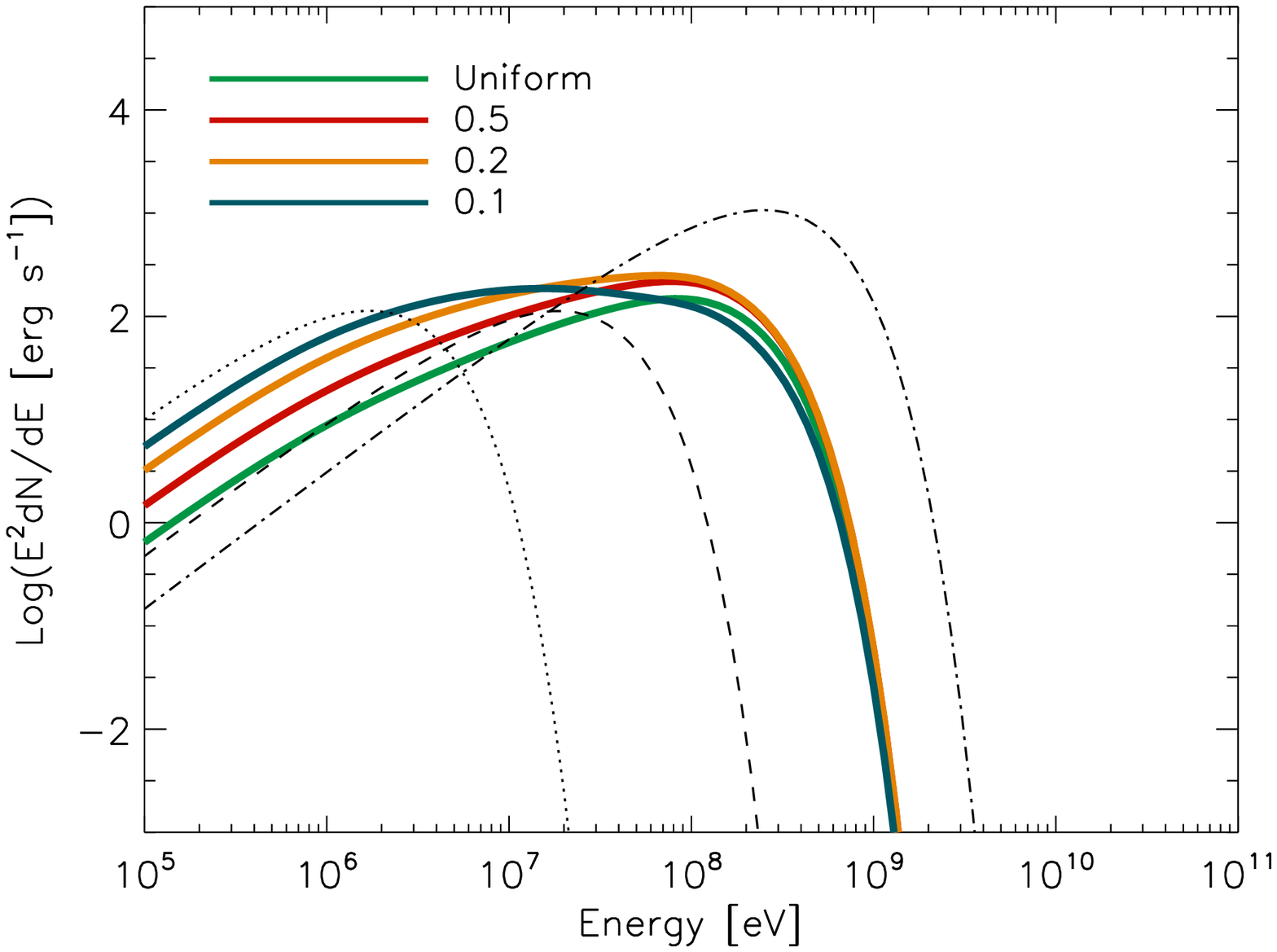}\\
\includegraphics[width=.52\textwidth]{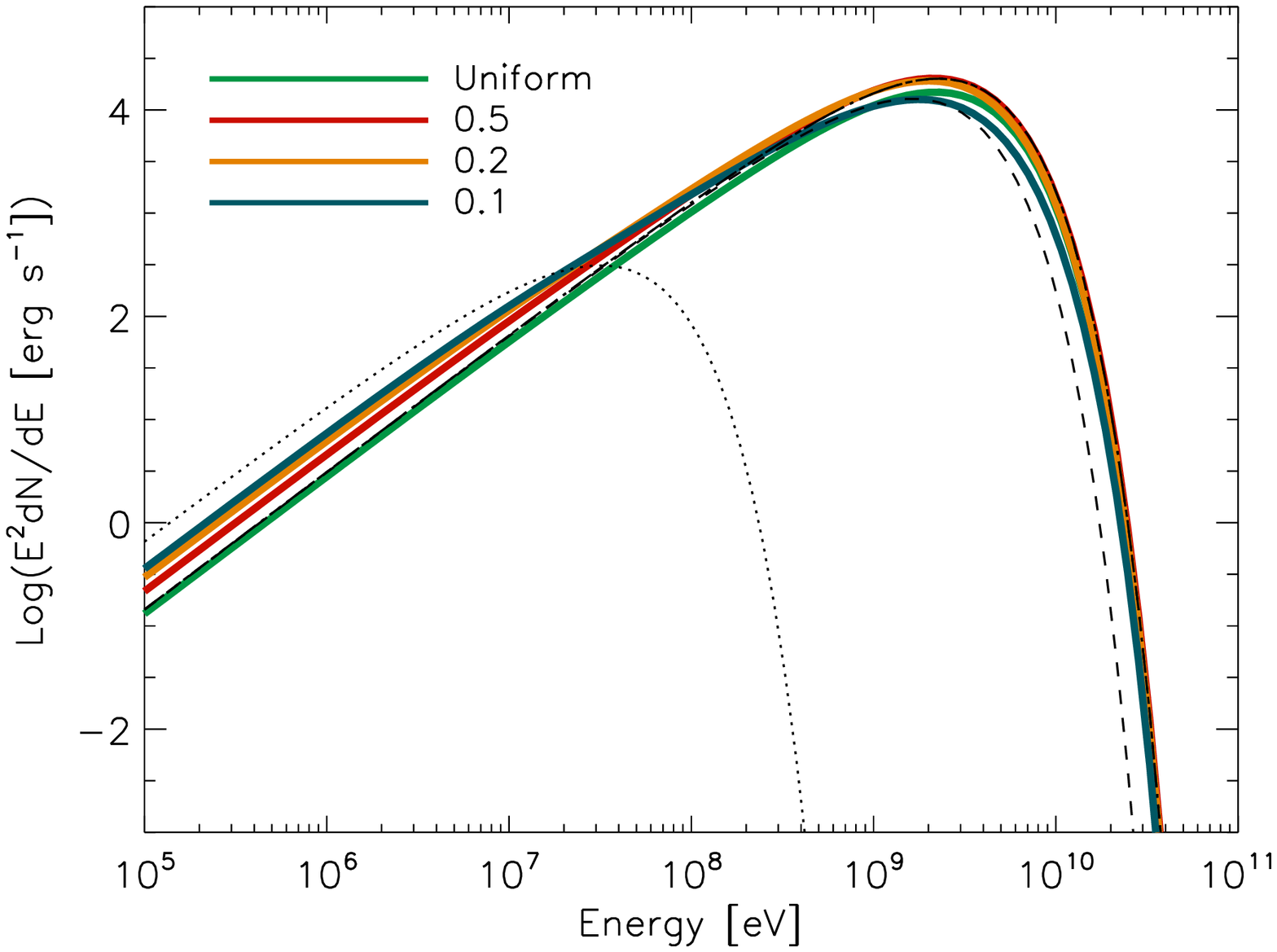}
\hspace{-1cm}
\includegraphics[width=.52\textwidth]{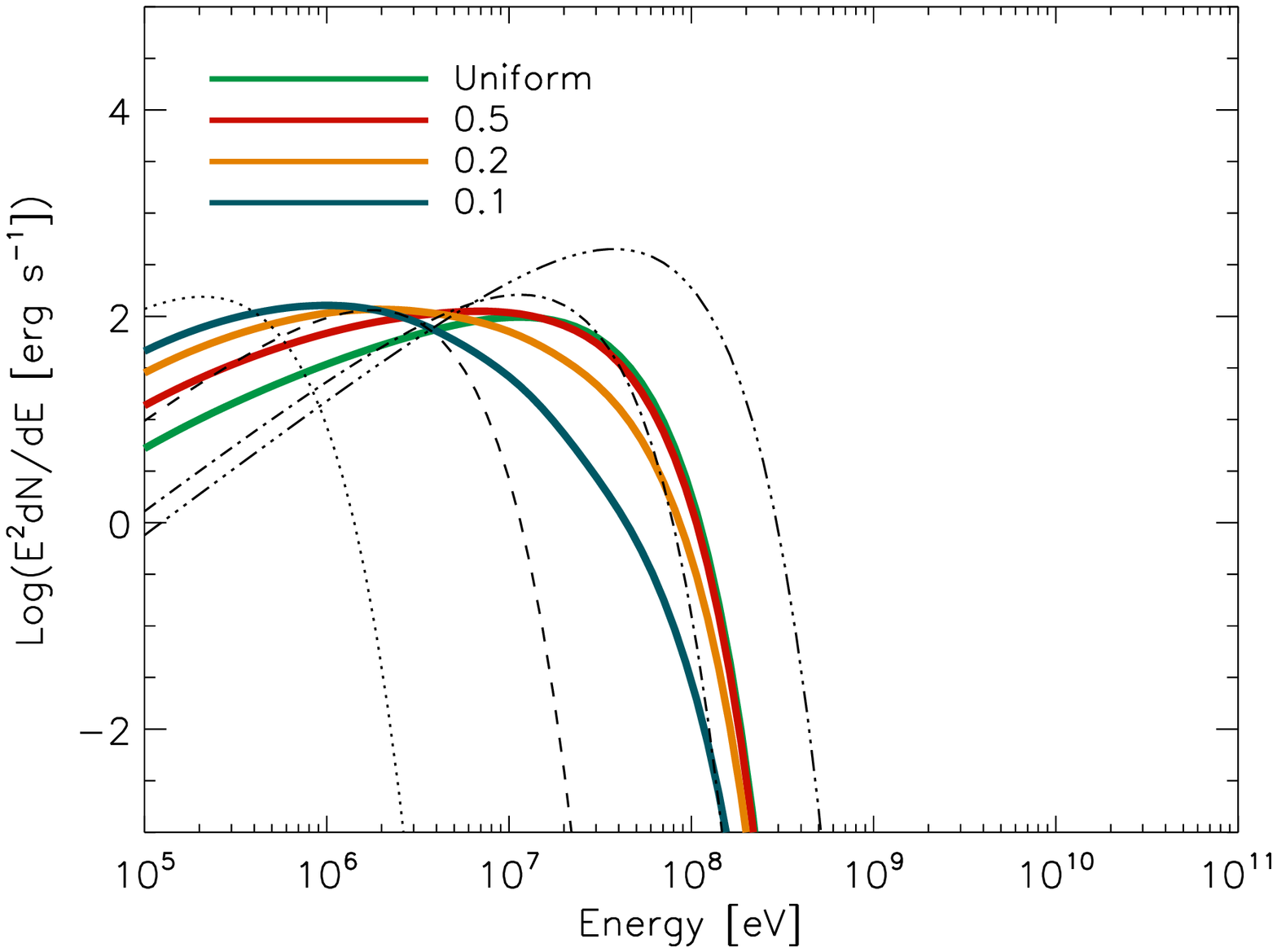}
\caption{Synchro-curvature spectra given by a single particle, for models A (top, left with $x_{\rm max}=r_c$, right with $x_{\rm max}=0.5r_c$), B (bottom left, $x_{\rm max}=1$), E (bottom right, $x_{\rm max}=1$). The dashed, dot-dashed, triple dot-dashed, and long-dashed lines represent the spectrum at a distance $x=0.01,0.1,0.5,1~r_c$, respectively. The solid lines represent the average spectra, weighted either uniformly, or with an exponential distribution with different widths (see text, values of $x_0/r_c$ are indicated in the legend).}
 \label{fig:spectra_tra2d}
\end{figure*}

\section{Synchro-curvature spectra}\label{sec:spectra}

The spectral energy distribution of photons emitted by a single particle is given by eq.~(\ref{eq:sed_synchrocurv}):
\begin{equation}\label{eq:sed}
 \frac{dP_{sc}}{dE} = \frac{\sqrt{3} e^2 \Gamma E}{4\pi \hbar r_{\rm eff} E_c}[(1 + z) F(y) - (1 - z) K_{2/3}(y) ]~,
\end{equation}
where we employ the factors given by eqs.~(\ref{eq:characteristic_energy}), (\ref{eq:q2_simple}), (\ref{eq:def_reff}), (\ref{eq:def_gr}), (\ref{eq:def_xi}) and (\ref{eq:def_rgyr}). Note that, for $E\ll E_c$, then $E^2dN/dE \propto E^{1.25}$, because, in that regime, $F(y) \sim y^{-0.75}$. Below, we explore the dependence of the spectrum on the particle position, and compute the spectrum integrated along the trajectory.  

We have just proven that at the beginning of the trajectory, $\xi$ is large, the Lorentz factor is relatively low, and the spectrum peaks at low energies. As $\Gamma$ grows under the electric acceleration, and, at the same time, the perpendicular momentum is dissipated, the spectrum approaches the purely curvature formula, and the spectrum peaks at higher energies.

In order to calculate the average spectrum radiated throughout the trajectory, we consider the spectra emitted at a grid of distances from zero (the place of particle injection) up to a maximum value $x_{\rm max}$. In all cases, we check that the spatial resolution was large enough to ensure numerical convergence. We want to explore the effects of giving more relevance to the high-$\xi$ parts of the trajectory. For this purpose, we calculate the average spectra as

\begin{equation}\label{eq:sed_wei}
 \frac{dP_{\rm gap}}{dE} = \int _{0}^{x_{\rm max}} \frac{dP_{sc}}{dE} w(x) dx~,
\end{equation}
where $w(x)$ is the relative weight of the contributions from different positions. We explore a uniform weight, $w(x)=1/x_{\rm max}$, and an exponential distribution (normalized to give $\int_0^{x_{\rm max}} w(x) dx =1$):
\begin{equation}\label{eq:weight}
w(x)=\frac{e^{-x/x_0}}{x_0(1-e^{-x_{\rm max}/x_0})}~.
\end{equation}
Below we set three different values of $x_0=0.5,0.2,0.1~r_c$. In the context of the outer gap, the different weights or the different $x_{\rm max}$ considered here could be qualitatively interpreted as the net effect of beaming effects, mentioned in \S\ref{sec:synchrocurvature}. Eqs.~(\ref{eq:beam1}) and (\ref{eq:beam2}) tell us that the beaming of the radiation covers a solid angle $\sim 1/\Gamma^2$.

First, depending on the viewing angle, the detectable radiation may be originated only along a part of the trajectory of the particles. In particular, the relatively small beam of the radiation coming from low-$\Gamma$ particles are more likely to be detected, unless the axis of the beam exactly matches the observer direction. Second, the solid angle is larger for the softer photons, giving a net softening effect for off-axis observer. 

A third possible effect is that particles that reverse its motion due to the sign of its charge are subject to strong synchrotron losses due to the necessary large 
perpendicular moment. These may be subdominant to the other effects but will sum up to a possible increase of the weight of the more internal regions. 

We remark that all these effects are hardly quantificable without detailed multidimensional numerical simulations including particle dynamics, geometry and beaming effects. However, we can qualitatively infer the combination of such effects would produce a systematically larger weight in the inner parts. The proposed functional form of the parametrization of the weight, Eq.~(\ref{eq:weight}), is not quantitatively justified and is only one of the possible arbitrary, phenomenological choices to effectively consider such possible beaming effects.

In Fig.~\ref{fig:spectra_tra2d} we show the spectra for the models A (top left, with $x_{\rm max}=r_c$; top right, with $x_{\rm max}=0.3 r_c$), B (bottom left, with $x_{\rm max}=r_c$) and E (bottom right, with $x_{\rm max}=r_c$).  The dashed, dot-dashed, triple dot-dashed, and long-dashed lines represent the spectrum at a distance $x=0.01,0.1,0.5,1~r_c$, respectively. The solid lines represent the uniform and weighted average spectra. At high energy, the total spectrum is well approximated by the curvature radiation emitted when $\sin\alpha \ll 1$ and $\Gamma = \Gamma_{\rm st}$. At lower energies, if the relative weight of the synchrotron-dominated emission is large enough, then the total spectrum can peak at lower energies than the curvature radiation spectra. As a consequence, the slope of the spectrum before the peak depends on the relative weight, the value of $x_{\rm max}$, and the model parameters. The smaller $x_{\rm max}$, the lower the peak of the spectrum, and the larger the relative weight of the high-$\xi$ parts of the trajectory. This effect could then lead to a variety of synchro-curvature spectra.

In model A with a smaller value of $x_{\rm max}$, and in model E, the spectra peak at lower energies because of the lower values of $\Gamma$. In the case of very strong electric field, a large value of $\Gamma$ is reached very fast, so that we always see a spectrum compatible with a purely curvature one. Of course, these models represent a trend, so that the combined variations of all parameters can lead to very different slopes and energy peaks.


\section{Astrophysical applications}\label{sec:discussion}

\begin{table*} 
\caption{Synchro-curvature parameters for different astrophysical scenarios. The radius of curvature is assumed to be of the order of the typical length-scale of the problem. Unknown values are marked by ??.}
\begin{tabular}{lccccc} 
\hline 
\hline
 & Lengthscale & $B$ & $\sin\alpha$ & $\Gamma$ & References \\
Scenario & [cm] & [G] & & & \\
\hline
 Solar flares & $10^7$-$10^{10}$ & 1-10 & 0.1? & 1 & \cite{hudson95} \\
 Pulsar gaps & $10^7$-$10^9$ & $10^4$-$10^7$ & $\ll 1$ & $10^7$-$10^8$ & \cite{cheng86a} \\
 Pulsar wind nebulae & $10^{18}$-$10^{20}$ & $10^{-6}$-$10^{-5}$ & ?? & $10^4$-$10^6$ & \cite{torres14} \\
 High Mass X-ray binaries & $10^{10}$ & $1$-$10$ & ?? & $10^2$ & \cite{dubus13} \\
 Transitional pulsars (propeller) & $10^6$-$10^7$ & $10^5$-$10^6$ & ?? & $10^4$ & \cite{papitto14} \\
\hline 
\end{tabular} 
\label{tab:scenarios} 
\end{table*} 


Despite the fact that synchro-curvature radiation has been mainly studied in the context of pulsar gaps, a variety of applications could require such formalism. In Table~\ref{tab:scenarios}, we indicate some astrophysical scenarios, evaluating the characteristic quantities with regards of synchro-curvature radiation. The radius of curvature of the magnetic field is taken to be of the order of the length-scale of the system. This is of course not true if the magnetic field is turbulent, or dominated by small scale structure (like in GRBs). The magnetic field can be measured directly, e.g. in galaxies, or indirectly, e.g., in pulsars by the timing properties of the pulsar plus an assumption on the location of the gap and the radial dependence of the magnetic field, or, in PWNe, by the magnetic energy required to explain the spectrum.
The Lorentz factor is inferred from the modeling of the observed spectra, assuming a dominant radiation mechanism from the multi-wavelength spectrum. Thus, while $r_c$ and $B$ can be independently (roughly) estimated or inferred, the evaluation of $\Gamma$ and $\alpha$ are intrinsically related to the dynamical and radiative assumed models, which depend on $\xi$.


Most of the astrophysical scenarios lie in the synchrotron regime, which justifies neglecting curvature radiation as an important mechanism. For instance, in the solar flares, particles are non-relativistic, which makes $\xi\gg1$: they emit weak, purely synchrotron radiation, which is detected as non-thermal X-ray radiation. In pulsar wind nebulae (PWNe), the inferred magnetic fields are of the order of tens of $\mu$G \citep{torres14}, but the Lorentz factors are, on average, lower than in the outer gap. Special conditions in nebulae with non-uniform magnetic fields can greatly affect this conclusion.

There is much uncertainty on the location where $\gamma$-rays are produced in the case of the high-mass X-ray binaries (HMXBs). One of the most promising mechanisms lies in the pulsar wind / stellar wind shock. In such case, the inverse Compton efficiently pushes the intense flux of low-energy photons originated from the massive star, to provide the high-energy part of the spectrum. Under this assumption, the synchrotron emission dominates in X-rays, $\Gamma\sim 10^2$, and the curvature radiation contribution is negligible, if the radius of curvature is of the order of the typical length-scale, $\sim 10^{12}$ cm, and $B\sim 1$ G \citep{dubus13}. In the context of low-mass transitional pulsars, the propeller model proposed by \cite{papitto14} for XSS~J12270-4859 provides $\Gamma\sim 10^4$, again assuming that the main mechanism of $\gamma$-ray radiation is the inverse Compton scattering on the background photons emitted in the disk and the synchrotron photons themselves. 
%
%

\section{Conclusions}

In this work we have revised the well-known synchro-curvature radiation formulae proposed by \cite{cheng96}. These formulae were initially derived in the context of the outer gap model for $\gamma$-ray pulsars. However, the problem of synchro-curvature radiation is very general,  applying to any astrophysical scenario where particles are accelerated. 

We have found a more compact formulation for the sychro-curvature power, which relies on the parameter $\xi$, which is related to the ratio between the Larmor and curvature radii. An evaluation of the typical magnetic field, radius of curvature, pitch angle, and Lorentz factor immediately allows to quantitatively estimate whether the synchrotron or the curvature radiation dominates, or if $\xi$ is of the order 1, the whole synchro-curvature formalism should be considered.  We reinforce the idea that the synchro-curvature radiation affect the spectrum and the particle motion, in the pulsar gaps.

We have numerically solved the particle dynamics for assumed values of electric field, radius of curvature, and magnetic field. We have explored the position-dependent, single-particle spectra, as well as the average spectra. Depending on the model parameters and the assumed weight of the different parts of the trajectories, the average spectrum can deviate considerably from a purely curvature spectrum, especially for the slope $\eta$ before the peak (the low-energy slope), $E^2dN/dE \propto E^{\eta}$. Since the synchro-curvature spectra predicts $\eta=1.25$ for a single particle at a specific position, the low-energy slope in the average spectra is always $\eta\leq 1.25$, and can be as low as $\eta\sim 0$ if most of the radiation we detect comes from the initial, high-$\xi$ part of the trajectory. On the other hand, an exponential cut-off at high energy is always predicted, as is due to the high-$\Gamma$ particles.

The study of the low-energy slope is of particular interest when comparing with the observed spectra \citep{2fpc}, which show $0\lesssim \eta \lesssim 1.25$. We argue that geometrical effects (viewing angles) and/or inhomogeneous distribution of the places of pair creation can qualitatively be reproduced by a weighted spectrum, which can lead to the observed range of low-energy slopes. A more detailed modeling of the outer gap of pulsars, and the expected spectra, can be found in \cite{paper2}.


\section*{Acknowledgements}

This research was supported by the grant AYA2012-39303 and SGR2014-1073 (DV, DFT). KH is partly supported by the Formosa Program between National Science Council in Taiwan and Consejo Superior de Investigaciones Cientificas in Spain administered through grant number
 NSC100-2923-M-007-001-MY3. 
The research leading to these results has also received funding from the
European Research Council under the European Union's Seventh Framework
Programme (FP/2007-2013) under ERC grant agreement 306614 (MEP). MEP
also acknowledges support from the Young Investigator Programme of the
Villum Foundation. We thank the referee for very useful comments and suggestions which have substantially improved the work.

\begin{figure}
\centering
\includegraphics[width=.4\textwidth]{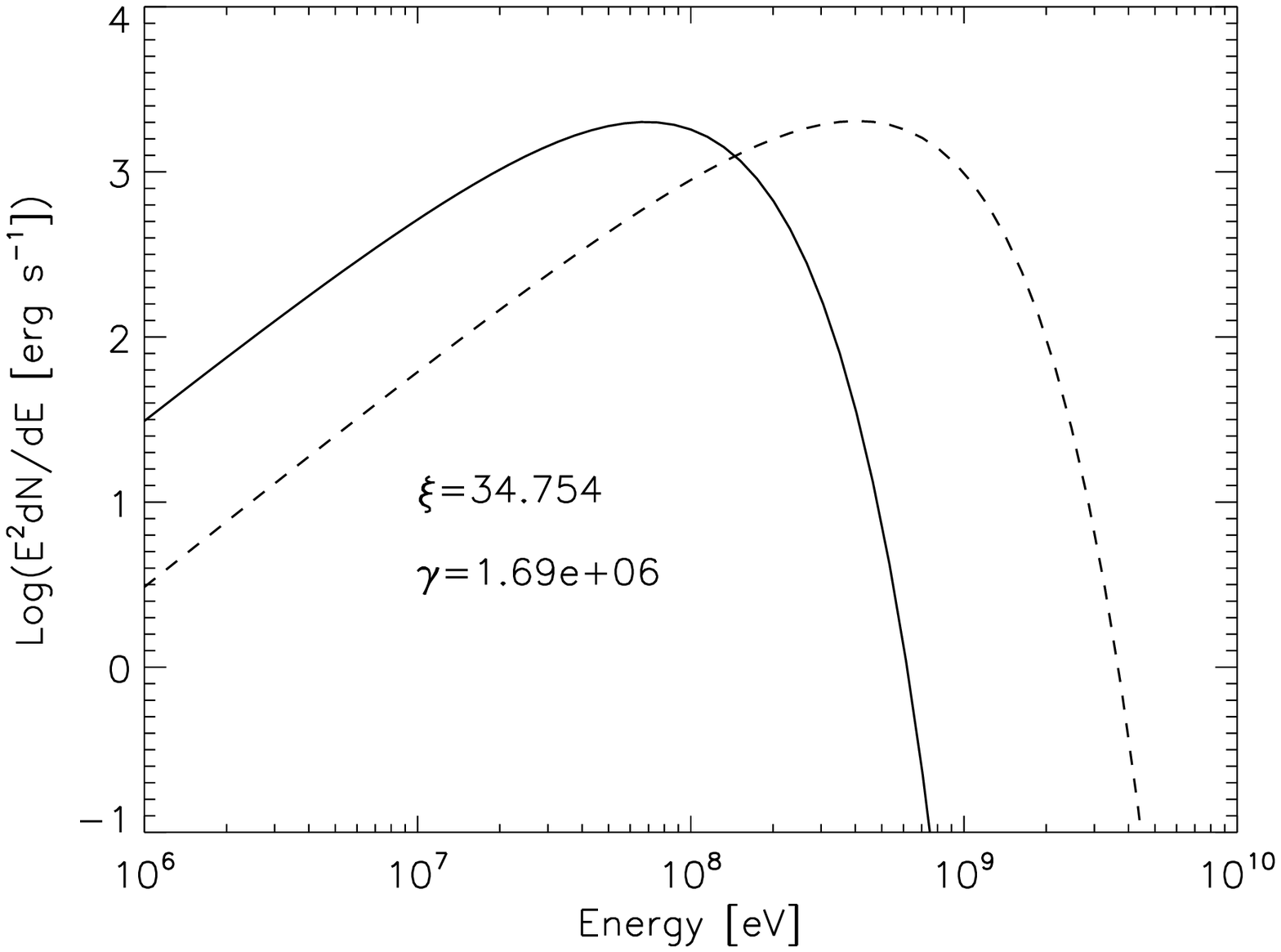}\\
\includegraphics[width=.4\textwidth]{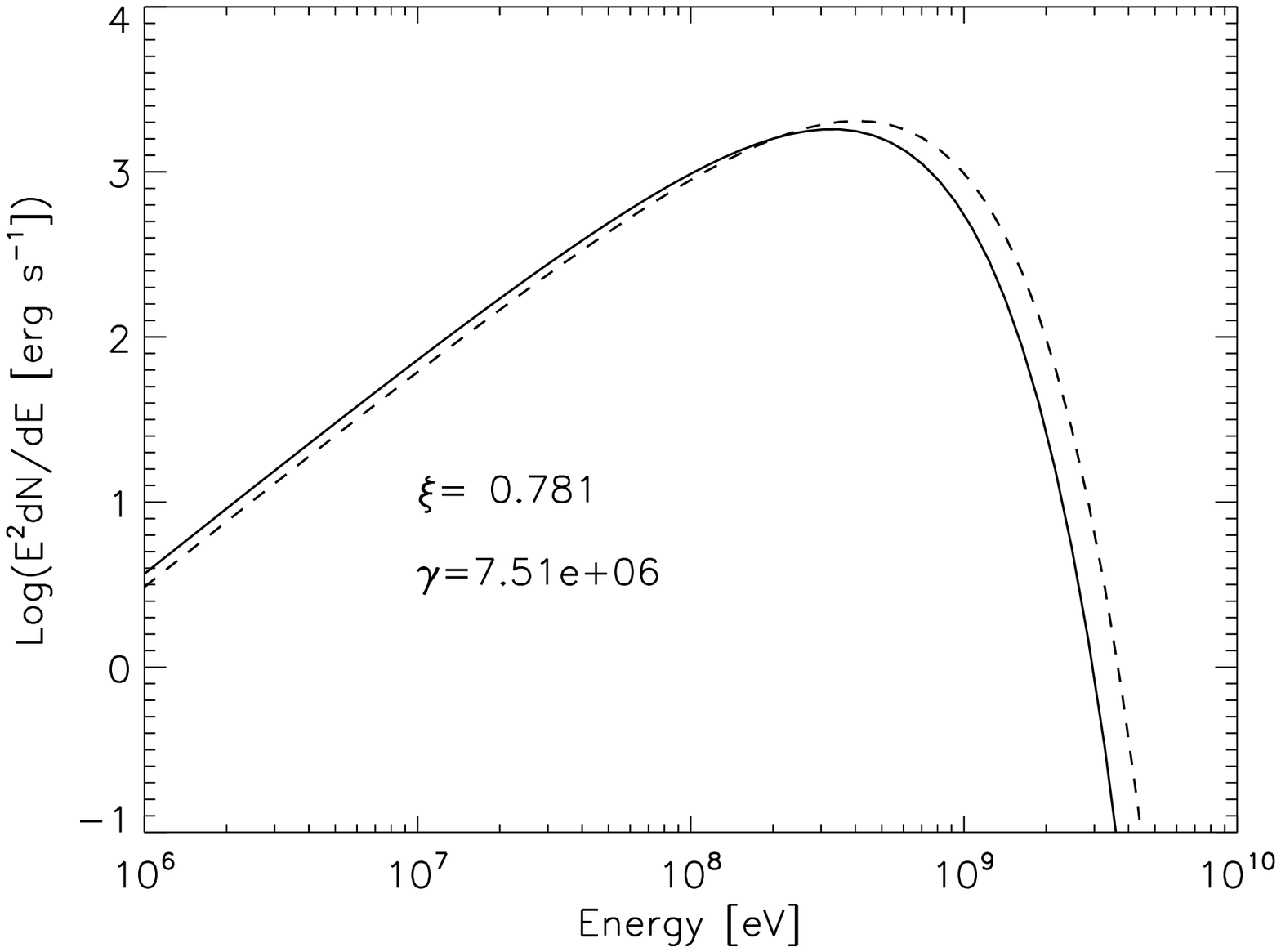}
\caption{Single particle synchro-curvature spectra (solid lines) for model A at saturation regime, with fixed $\sin\alpha=10^{-3}$ (top) and $\sin\alpha=10^{-4}$ (bottom). Dashed lines show, for comparison, the purely curvature radiation. We indicate in each panel the saturation values of $\Gamma$ and $\xi$.}
 \label{fig:spectra_single}
\end{figure}

\section*{Appendix: Spectra for fixed pitch angle}

Let us now suppose that $\sin\alpha$ is fixed at some value, as considered in the thick outer gap model by \cite{zhang97}. In that case, the particles will reach a saturation value which is regulated by the synchro-curvature emission, eq.~(\ref{eq:motion_par}), instead of purely curvature one. In this case, one has to solve the system of equations reported altogether below:
\begin{eqnarray}
 && \Gamma = \left( \frac{3}{2}\frac{E_{\parallel} r_c^2}{e g_r} \right)^{1/4}~, \label{eq:gamma}\\
 && E_c = \frac{3}{2}\hbar cQ_2\Gamma^3~,  \label{eq:ec}\\
 && r_{\rm gyr}=\frac{mc^2\Gamma\sin\alpha}{eB}~,\\
 && \xi = \frac{r_c}{r_{\rm gyr}}\frac{\sin^2\alpha}{\cos^2\alpha}~,\\
 && Q_2 = \frac{\cos^2\alpha}{r_c}\sqrt{1 + 3\xi  + \xi^2 + \frac{r_{\rm gyr}}{r_c}} \label{eq:q2}~, \\
 && r_{\rm eff} = \frac{r_c}{\cos^2\alpha}\left(1 + \xi+ \frac{r_{\rm gyr}}{r_c}  \right)^{-1}~,\\
 && g_r =  \frac{r_c^2}{r_{\rm eff}^2}\frac{[1 + 7(r_{\rm eff}Q_2)^{-2}]}{8 (Q_2r_{\rm eff})^{-1}}~.\label{eq:gr}
\end{eqnarray}
If we fix a set of values for $B$, $r_c$, $e$, $m$, $E_\parallel$, $\alpha$, solving the system of coupled non-linear equations (\ref{eq:gamma})-(\ref{eq:gr}) corresponds to finding the fixed point of a function of $\Gamma$ (which is trivial only in the limits of curvature/synchrotron regimes), i.e. $\Gamma=f(\Gamma,B,r_c,e,m,E_\parallel,\alpha_{\rm in},\Gamma_{\rm in})$. We solve the system with a simple fixed point iteration method: we start with an initial guess for $\Gamma$, which is used to evaluate eqs.~(\ref{eq:ec})-(\ref{eq:gr}). The obtained value of $g_r$ is used to correct the value of $\Gamma$, eq.~(\ref{eq:gamma}). The iterative procedure is repeated until the values of $\Gamma$ numerically converge. We have verified that the same set of values is recovered for a large range of initial guess values, $\Gamma_{\rm g}\sim 10^4-10^9$. In all cases, the numerical convergence is achieved within maximum $\sim 10$ cycles.

In Fig.~\ref{fig:spectra_single} we show the synchro-curvature spectra for a single particle, for model A, fixing $\sin\alpha=10^{-3}$ (top) or $\sin\alpha=10^{-4}$ (bottom). The differences between the purely curvature (dashes) and the synchro-curvature (solid lines) increase with $\xi$. For large $\xi$, the loss of high energy photons is compensated by a larger flux at lower energies. The slope before and after the peak, however, does not change.

\bibliography{og}

\end{document}